\documentclass{jors}
\usepackage[square,numbers]{natbib}
\usepackage{har2nat}
\usepackage{hyperref}
\usepackage{breakurl}
\usepackage{graphicx}
\usepackage{amsmath}

\definecolor{mygray}{gray}{0.6}

\begin{document}

{\bf Software paper for submission to the Journal of Open Research Software} \\

To complete this template, please replace the blue text with your own. The paper has three main sections: (1) Overview; (2) Availability; (3) Reuse potential. \\

Please submit the completed paper to: editor.jors@ubiquitypress.com

\rule{\textwidth}{1pt}

\section*{(1) Overview}

\vspace{0.5cm}

\section*{Title}

Extended Dynamical Causal Modelling for Phase Coupling (eDCM PC)

\section*{Paper Authors}

1. Yeldesbay, Azamat; \\
2. Daun, Silvia; 

\section*{Paper Author Roles and Affiliations}

1. Role: designing, implementing and testing the program code, writing the article and the documentation, creation of the figures. \\
Affiliation: \\
(a) Institute of Zoology, University of Cologne, Cologne, Germany \\
(b) Research Centre J{\"u}lich, Institute of Neuroscience and Medicine, Cognitive Neuroscience (INM-3), 52425 J{\"u}lich, Germany

2. Role: Supervision, writing the article and the documentation, revising the text and the figures. \\
Affiliation: \\
(a) Research Centre J{\"u}lich, Institute of Neuroscience and Medicine, Cognitive Neuroscience (INM-3), 52425 J{\"u}lich, Germany \\
(b) Institute of Zoology, University of Cologne, Cologne, Germany

\section*{Abstract}

We present a software tool - extended Dynamic Causal Modelling for Phase Coupling (eDCM PC) - that is able to estimate effective connectivity between any kind of oscillating systems, e.g. distant brain regions, using the phase information obtained from experimental signals. With the help of a transformation function eDCM PC can measure observable independent coupling functions within and between different frequency bands. 

eDCM PC is written in the numerical computing language MATLAB as an extension to Dynamic Causal Modelling (DCM) for phase coupling (Penny et al. 2009)\cite{Penny2009}. eDCM PC is available on GitLab under the GNU General Public License (Version 3 or later).

\section*{Keywords}

phase oscillators; coupling functions; phase reduction; electroencephalography (EEG); magnetoencephalography (MEG); oscillatory signals

\section*{Introduction}

An oscillatory process is a wide spread phenomenon in nature that occurs in physical and biological systems  \cite{Winfree1980,Pikovsky2001,Strogatz2004}. Oscillatory and rhythmic activities play a prominent role in the interaction between biological systems, especially in the communication between brain areas \cite{Buzsaki2006}. Therefore it is of particular interest to understand how the interaction between the elements of these oscillating systems, e.g. between distant brain regions, occurs. In this context phase reduction is a useful method to analyze a large oscillating network by representing every oscillating system with one variable - the phase. 

\begin{figure}[h]
 \centering
 \includegraphics[width=0.7\textwidth]{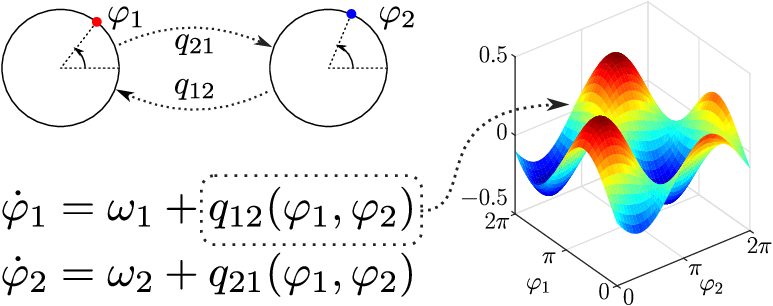}
 \caption{The model of two weakly coupled phase oscillators. A coupling function can be represented as a surface.\label{fig:wcpo}}
\end{figure}

In the theory of synchronization the interaction between oscillating systems is analyzed by the model of weakly coupled phase oscillators \cite{Pikovsky2001,Kuramoto1984,Hoppensteadt1997}:
\begin{equation}\label{eq:wcpo}
\dot{\varphi}_i = \omega_i + Q_i(\varphi_1,\ldots,\varphi_N), \; i=1,\ldots,N,
\end{equation}
where $\varphi_i$ is the phase of an oscillating system (an oscillator) $i$, $\omega_i$ is the natural frequency of the oscillator $i$, $Q_i$ is the interaction function (coupling function) with other oscillators. For weak coupling the coupling function $Q_i$ in Eq.~(\ref{eq:wcpo}) can be simplified as the sum of the pairwise coupling functions $q_{i,j}$:
\begin{equation}\label{eq:wcpo_q}
\dot{\varphi}_i = \omega_i + \sum_{j=1, j\neq i}^{N}q_{i,j}(\varphi_i,\varphi_j).
\end{equation}
The pairwise coupling functions $q_{i,j}$, on the other hand, can be represented as 2D surfaces (Fig.~\ref{fig:wcpo}).

One can obtain the coupling between the phase oscillators in the model Eq.~(\ref{eq:wcpo_q}) by finding the coupling functions $q_{i,j}$ directly from the experimental data \cite{Rosenblum2001}. This method was actively developed in the past decades and used to find the directionality of the couplings, the causal relations and to build dynamical models \cite{Kralemann2007,Kralemann2008,Kralemann2011,Stankovski2012,Kralemann2013,Kralemann2014,Stankovski2015,Stankovski2017,Pikovsky2018,Stankovski2019}. 

\begin{figure}[h]
 \centering
 \includegraphics[width=\textwidth]{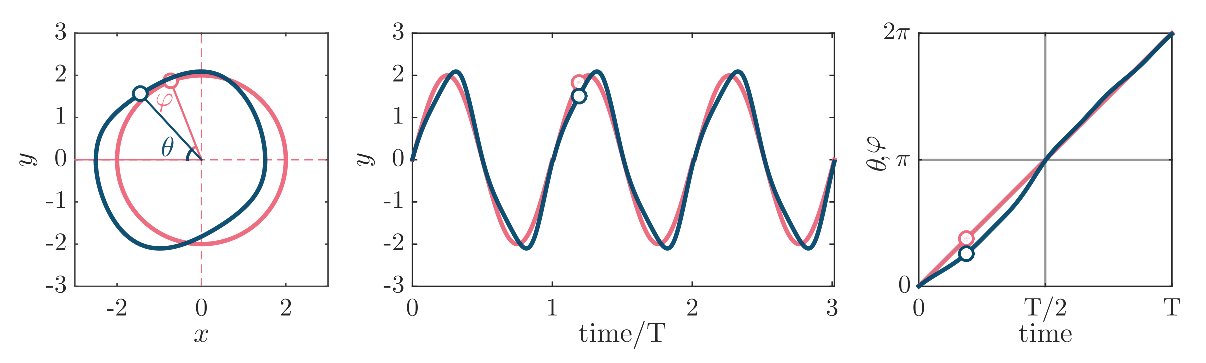}
 \caption{The difference between the theoretical and the observable (measured) phase when the limit cycle of the oscillating system is not circular (left panel). The signal of the theoretical phase is a sine wave, whereas the one of the observable phase is not (center panel). The observable phase grows non-linearly (right panel).\label{fig:thvsobs}}
\end{figure}

Kralemann and colleagues \cite{Kralemann2007,Kralemann2008} have shown that the phase extracted from the experimental signals (e.g. by using Hilbert or Wavelet transformation) behaves differently than the phase described in the theoretical model Eq.~(\ref{eq:wcpo}). The difference appears if the limit cycle of the oscillating system, the signal which we are measuring, is not circular, and the oscillatory signal has a non-sinusoidal form. This leads to a non-linear growth of the measured phase, even if the oscillating system has no input, as demonstrated in Fig.~\ref{fig:thvsobs}. Moreover, it was shown in \cite{Kralemann2008} that this non-linear growth of the measured phase can cause spurious couplings in the system under investigation. The problem was resolved by introducing a transformation function between the observable and the theoretical phases (referred to as the proto- and the true phases in \cite{Kralemann2008}). Kralemann and colleagues implemented this approach into their DAMOCO toolbox \footnote{The Data Analysis with Models Of Coupled Oscillators (DAMOCO) toolbox can be found here http://www.stat.physik.uni-potsdam.de/\textasciitilde mros/damoco.html} by approximating the transformation function from the observable phase to the theoretical phase using the distribution of the observable phase over a long period of time. 

The software tool presented in this metapaper - extended Dynamical Causal Modelling for Phase Coupling (eDCM PC) - addresses the problem of a potential non-linear growth in the observable phase. It is implemented as an extension to Dynamical Causal Modelling for phase coupling (DCM PC) \cite{Penny2009}. eDCM PC aims at finding the effective coupling in a network of oscillating systems by inferring the coupling function from the measured signals. In particular, the software tool is designed to reconstruct the coupling functions for the cases when the phase is not uniformly distributed, i.e. in the case when the transformation between the observable and the theoretical phases should be taken into account. Moreover, eDCM PC extends DCM PC by allowing to find the coupling between different frequency bands, thereby making it possible to analyze n:m synchronization cases. In contrast to the DAMOCO toolbox, eDCM PC uncovers the transformation functions from the theoretical to the observable phases together with the coupling functions. It uses the capability of DCM, based on Bayesian inference, to analyze and compare several possible network structures at the same time. 

The theoretical results related to eDCM PC and the numerical testing on synthetic data sets were presented in our previous work \cite{Yeldesbay2019}. Since its first presentation in \cite{Yeldesbay2019} eDCM PC was supplemented with a documentation, additional tests and plotting functions for data sets and results, and was rearranged in a user-friendly structure. In this work we give a detailed description on the architecture and usage of eDCM PC, as well as provide testing examples. 

\section*{Implementation and architecture}

The architecture of eDCM PC is presented in Fig.~\ref{fig:eDCMPCsch} as a data flow with different interface levels: data preprocessing level, user-interface level, eDCM PC level, and SPM/DCM level. We describe the levels in bottom up direction, namely starting from the SPM/DCM level up to the eDCM PC level by indicating specific implementation details. Thereafter, in the usage part, we describe the user-interface level. 

\subsubsection*{Dynamic causal modelling}

The structure of the eDCM PC is imposed by the architecture of the Dynamic Causal Modelling (DCM) \cite{Friston2003}. DCM is an opensource toolbox within the Statistical Parametric Mapping (SPM) software \cite{Friston2007}, developed to analyze connectivity in the brain. DCM can work with different modalities (functional magnetic resonance imaging (fMRI), EEG, MEG, local field potentials (LFP), functional near-infrared spectroscopy (fNIRS)) \cite{Friston2007b,Friston2003,David2006,Chen2008,Chen2012,Stephan2008,Moran2009,Tak2015,Stephan2010,Daunizeau2011}. The usage of a wide variety of modalities is provided by a particular feature of the DCM architecture that allows the modification of its different parts without changing the common, basic structure of the software \cite{Daunizeau2011}. 

\begin{figure}[h!]
 \includegraphics{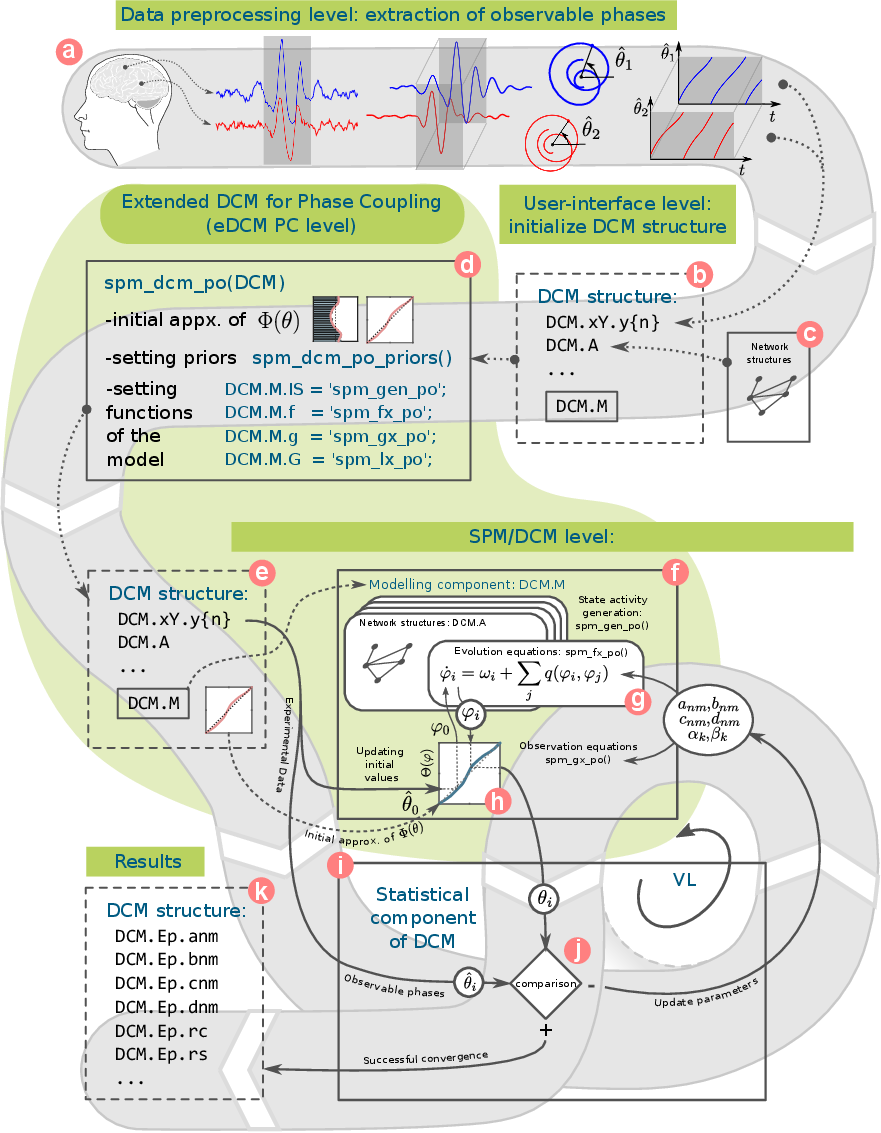}
 \caption{The data flow and the architecture scheme of eDCM PC. The green background area denotes the elements introduced in eDCM PC. \label{fig:eDCMPCsch}}
\end{figure}

DCM uses Bayesian inference to find the parameters of the system and the coupling between the brain regions. The elements of the common structure of DCM are the data preprocessing component (Fig.~\ref{fig:eDCMPCsch},\textbf{a}), the modelling component (Fig.~\ref{fig:eDCMPCsch},\textbf{f}), and the statistical component (Fig.~\ref{fig:eDCMPCsch},\textbf{i}). These elements can be modified with respect to the modality and the modelling, however the interaction between them remains the same for all versions of DCM. 

Initial raw data are transformed into time courses of observables, e.g. into observable phases $\hat{\theta}$ (Fig.~\ref{fig:eDCMPCsch},\textbf{a}). In the modelling component (Fig.~\ref{fig:eDCMPCsch},\textbf{f}) synthetic signals are generated by numerically integrating the model with an initial guess for the system parameters (priors). The modelling component consists of the evolution equations (Fig.~\ref{fig:eDCMPCsch},\textbf{g}), a system of ODEs that represents the hidden state of the system and the connections between parts of the system, and the observation equations (Fig.~\ref{fig:eDCMPCsch},\textbf{h}), functions that express the observables using the hidden states. In the statistical component of DCM (Fig.~\ref{fig:eDCMPCsch},\textbf{i}) the observables $\hat{\theta}$ obtained from the data and the observables $\theta$ from the synthetic signals are "compared" by means of the Bayesian model inversion, namely Variational Laplace (VL) (Fig.~\ref{fig:eDCMPCsch},\textbf{j}). In a simplified manner, the result of this comparison gives the correction for the system parameters. Using these corrected parameter values the modelling component generates new synthetic signals. This procedure repeats until the required convergence between the measured and the synthetically generated observables is reached (Fig.~\ref{fig:eDCMPCsch},\textbf{k}). 

\subsubsection*{Modelling component}
eDCM PC introduces a new modelling component. As mentioned before the modelling component contains the evolution equation and the observation equation. The extended evolution equation is defined as
\begin{equation}\label{eq:evol}
\dot{\varphi}_i = \omega_i + \sum_{j=1, j\neq i}^{N_q} q_{ij}(\varphi_i,\varphi_j),
\end{equation}
where the pairwise coupling functions are constructed in the way to detect $`n:m`$ synchronization 
\begin{equation}\label{eq:qij}
q_{ij}(\varphi_i,\varphi_j) = \sum_{n,m = -N_q, \; n,m\neq 0}^{N_q} Q_{ij}e^{i(n\varphi_i + m \varphi_j)},
\end{equation}
with complex Fourier coefficients $Q_{ij}$ truncated at a given number of terms $N_q$. In the code, however, the coupling functions $q_{ij}$ are represented 
with real-valued Fourier coefficients as
\begin{align}\label{eq:qijreal}
 q_{ij}(\varphi_i,\varphi_j) = \sum_{n=1}^{N_q}\sum_{m=1}^{N_q} \big[& a^{(n,m)}_{ij}
\cos(n\varphi_i)\cos(m\varphi_j) +  b^{(n,m)}_{ij}
\cos(n\varphi_i)\sin(m\varphi_j) \\
 &+c^{(n,m)}_{ij} \sin(n\varphi_i)\cos(m\varphi_j) + d^{(n,m)}_{ij}
\sin(n\varphi_i)\sin(m\varphi_j)\big].
\end{align}

To introduce the transformation function between the theoretical and observable phases (the forward transformation) we use the observation equation. Following \cite{Kralemann2008} we define the transformation as follows
\begin{equation}\label{eq:frwtr}
\theta_i=\Theta(\varphi_i) = \varphi_i + \sum_{k=-N_{\rho}, k\neq 0}^{N_{\rho}}(e^{ik\varphi_i}-1),
\end{equation}
also truncated at a given number of terms $N_{\rho}$. The forward transformation function is represented in the code with real-valued coefficients: 
\begin{equation}
  \theta_i = \Theta(\varphi_i) = \varphi_i +
\sum\limits_{k=1}^{N_{\rho}}\frac{1}{k}\left[\alpha_i^{(k)}\sin(n\varphi_i) -
\beta_i^{(k)}\cos(n\varphi_i) + \beta_i^{(k)}\right].
\end{equation}

Thus, in eDCM PC the hidden states are the theoretical phases $\varphi_i$ and the observables are the observable phases $\theta_i$ defined for every region and every frequency band. 

The resulting reconstructed coupling and transformation functions are (1) the real-valued coefficients (matrices) $a_{ij}^{(n,m)}$, $b_{ij}^{(n,m)}$, $c_{ij}^{(n,m)}$, $d_{ij}^{(n,m)}$ defined for every connected pair of regions and (2) the real-valued coefficients (vectors) $\alpha_i^{(k)}$ and $\beta_i^{(k)}$ defined for every region, respectively. 

\subsubsection*{Initial approximation of the observation equation}

\begin{figure}[h]
    \centering
    \includegraphics[width=\textwidth]{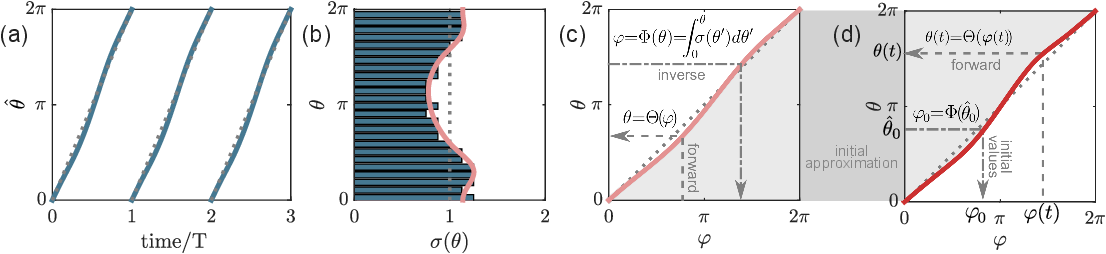}
    \caption{Initial approximation of the inverse transformation function and updating the initial conditions. \label{fig:appxinit}}
\end{figure}

Good convergence is provided by a good initial guess of the parameters of the system. Therefore, the initial approximation of the transformation function, i.e. the observation equation, is essential for a successful reconstruction of the system parameters and the coupling functions. eDCM PC uses the approach presented in \cite{Kralemann2008} to approximate the inverse transformation function $\varphi=\Phi(\theta)$, as shown in Fig.~\ref{fig:appxinit}, panels (a),(b), and (c).

The average distribution of the observable phases $\hat{\theta}_i$ over a long period of time for a region $i$ can be approximated by a $2\pi$ periodic function $\sigma_i(\theta)$ with the mean equal to 1. (Fig.~\ref{fig:appxinit}, (a) and (b)), and be written as the following Fourier series \cite{Kralemann2008}

\begin{equation}
 \sigma_i(\theta) = 1 + \sum_{k=1}^{N_{\sigma}} \hat{\alpha}_i^{(k)} \cos(k \varphi_i) + \hat{\beta}_i^{(k)} \sin(k \varphi_i),
\end{equation}
where the sum is truncated at $N_{\sigma}$. The integral of $\sigma_i(\theta)$ over one period gives us the inverse transformation function  
\begin{equation}\label{eq:invtrans}
 \Phi_i(\theta) = \theta_i +
\sum\limits_{k=1}^{N_{\sigma}}\frac{1}{k}\left[\hat{\alpha}_i^{(k)}\sin(k\theta_i) -
\hat{\beta}_i^{(k)}\cos(k\varphi_i) + \hat{\beta}_i^{(k)}\right],
\end{equation}

which is the inverse function of the forward transformation function $\Theta(\varphi)$ defined in Eq.~(\ref{eq:frwtr}) (Fig.~\ref{fig:appxinit}, (c)). Note, that the truncation orders $N_{\sigma}$ and $N_{\rho}$ are, in general, different. 

After obtaining the coefficients $\hat{\alpha}$ and $\hat{\beta}$ by approximating the average distribution of the observable phases $\sigma_i(\theta)$, eDCM PC finds the initial approximation of the coefficients $\alpha$ and $\beta$ of the forward transformation $\Theta(\varphi)$.

\subsubsection*{Updating the initial conditions}
In every step of the Bayesian estimation the updated parameters $\alpha$ and $\beta$ change the transformation function $\Theta(\varphi)$. Therefore the initial conditions $\hat{\theta}_0$ of the observations should also be recalculated in every step to obtain new initial conditions for the evolution equation $\varphi_0$ (Fig.~\ref{fig:appxinit}, (d)). For this the updated inverse transformation function $\Phi(\theta)$ is needed. Therefore, eDCM PC recalculates the coefficients $\hat{\alpha}$ and $\hat{\beta}$ of Eq.~(\ref{eq:invtrans}) using the updated coefficients $\alpha$ and $\beta$ after every step of the Bayesian estimation. 

\subsubsection*{Usage}

The data flow of the eDCM PC is presented in Fig.~\ref{fig:eDCMPCsch}. The eDCM PC package works with time courses of the observable phases obtained from oscillatory signals. The observable phase can be extracted using the Hilbert or Wavelet transformation applied to the raw signals filtered within a specific frequency band (Fig.~\ref{fig:eDCMPCsch}, (a)). 

In the user-interface level the usage of eDCM PC comes down to the initialization of the elements of the \verb|DCM| structure (Fig.~\ref{fig:eDCMPCsch}, (b)) and calling the \verb|spm_dcm_po(DCM)| function (Fig.~\ref{fig:eDCMPCsch}, (d)). 

The code of eDCM PC contains an example script \verb|test.m| that runs the reconstruction procedure using a pre-simulated data set. For details of the \verb|DCM| structure and calling the \verb|spm_dcm_po(DCM)| function we refer to the documentation included in the repository of eDCM PC. Here, we discuss only essential parts of the \verb|DCM| structure. 

The observable phases $\hat{\theta}$ should be stored in the \verb|DCM| structure in the fields \verb|DCM.xY.y{n}|, where \verb|n| is the trial index (Fig.~\ref{fig:eDCMPCsch}, (b)). All trials should have the same length with an equally distributed time step. 

The supposed network structure of the system is defined in the binary adjacency matrix \verb|DCM.A| (Fig.~\ref{fig:eDCMPCsch}, (c)). DCM allows the definition of several possible network structures and finds the structure best fitting the data using Bayesian model comparison. In that case \verb|DCM.A| is an array of matrix cells, and different structures should be defined in different cells. For details of the definition of the structure, please refer to the documentation of DCM in Chapters 41,42, and 43 of \cite{Friston2007}.

The elements of the modelling component are defined in the substructure \verb|DCM.M|, such as the mean frequency $\omega_i$ (\verb|DCM.M.freq|) and the frequency band (\verb|DCM.M.fb|) for every region, the truncation orders of the Fourier series for the coupling functions $N_{q}$ (\verb|DCM.M.Nq|), for the forward transformation functions $N_{\rho}$ (\verb|DCM.M.Nrho|), and for the inverse transformation functions $N_{\sigma}$ (\verb|DCM.M.Nsig|). 

The initialized \verb|DCM| structure is used as an argument to call the \verb|spm_dcm_po(DCM)| function (Fig.~\ref{fig:eDCMPCsch}, (d)). This function performs the initial approximation of the inverse transformation function $\Phi(\theta)$, defines the priors for all parameters of the system (\verb|spm_dcm_po_priors|), and assigns the functions of the model in the substructure \verb|DCM.M| such as:
\begin{itemize}
    \item state activities generation function \verb|spm_gen_po|;
    \item evolution equation function \verb|spm_fx_po|;
    \item observable equation function \verb|spm_gx_po|;
    \item linear observation function \verb|spm_lx_po|, which is a dummy function for compatibility with other versions of DCM. The function returns the values without change.
\end{itemize}
Thereafter, \verb|spm_dcm_po(DCM)| calls DCM routines with the \verb|DCM| structure as the input argument to start the Bayesian estimation of the parameters of the system (Fig.~\ref{fig:eDCMPCsch}, (e)). 

The result of the \verb|spm_dcm_po(DCM)| function is an extended \verb|DCM| structure that contains the estimated values of the system parameters in the substructure \verb|DCM.Ep| (Fig.~\ref{fig:eDCMPCsch}, (k)), namely the matrices of the Fourier coefficients of the coupling functions $a_{ij}^{(n,m)}$, $b_{ij}^{(n,m)}$, $c_{ij}^{(n,m)}$, $d_{ij}^{(n,m)}$ in \verb|DCM.Ep.anm|, \verb|DCM.Ep.bnm|, \verb|DCM.Ep.cnm|, \verb|DCM.Ep.dnm| respectively, of the forward transformations in \verb|DCM.Ep.rc| and \verb|DCM.Ep.rs|, which correspond to $\alpha_i^{(k)}$ and $\beta_i^{(k)}$ respectively. 

\subsubsection*{Installing and testing} 

eDCM PC is written in MATLAB and uses Dynamic Causal Modelling (DCM), which is included in the Statistical Parametric Mapping (SPM) package \cite{David2007}. The SPM package can be downloaded here \url{https://www.fil.ion.ucl.ac.uk/spm/software/download/}. The minimum version of SPM used by eDCM PC is SPM12. 

eDCM PC can either be downloaded or cloned from our GitLab repository \url{https://gitlab.com/azayeld/edcmpc}. 

For installation follow the instruction given in the documentation of the eDCM PC package.
After installation, eDCM PC can be tested by calling the \verb|test.m| script.   

\section*{Quality control}

The users can easily test whether the installation was correct by calling a \verb|test.m| script. 

The package includes two other example scripts, which are also intended to test the quality of the reconstruction. 

In the first example, eDCM PC successfully reconstructs the coupling and the transformation functions from the signals of the synthetically simulated system of two weakly coupled phase oscillators. In this example the parameters of the coupling are chosen such that one coupling is zero, i.e. making the coupling unidirectional. Moreover, an artificial distortion (transformation) is added to the theoretical phases in order to simulate the effect of non-linear growing of the observable phase. Thus, the parameter values reconstructed by eDCM PC can be compared with the original parameter values used by the simulation.

The second example analyzes two uni-directionally coupled neural mass models, namely, Jansen and Rit models, which simulate EEG-like signals. In this case neither the coupling function nor the non-linear relation between the observable and the theoretical phases are known, which simulates the real case scenario. In this example, eDCM PC demonstrates successful reconstruction of the coupling functions and shows the presence of a non-linear relation between the observable and theoretical phases in the neural mass models. 

The eDCM PC package includes several routines which allow testing of the data set before running the reconstruction procedure and examining the quality of the reconstruction afterwards:
\begin{enumerate}
 \item Routines that allow the analysis of the spectral relation between two regions by plotting an approximated 2D Fourier spectrum of the observable phases. These routines can be used to analyze the data before the reconstruction procedure to set the truncation orders of the Fourier series (\verb|DCM.M.Nq| and \verb|DCM.M.Nrho|). 
 \item Routines to plot the distribution of the observable phases together with the initial approximation of the inverse transformation function and the final reconstructed inverse transformation function. These routines can be used to set the value of the parameter \verb|DCM.M.Nsig|, and test the final reconstructed inverse function. 
 \item Routines to plot the reconstructed pairwise coupling functions together with the measured data points. These functions are useful to examine the convergence after the reconstruction procedure. Since the coupling functions $q_{ij}(\varphi_i,\varphi_j)$ and the measured data points (the measured observable phases $\hat{\theta}_i$ and $\hat{\theta}_j$) are in different domains, there are two possibilities to compare them: (1) by representing the data points on the domain of the theoretical phases $\varphi$ using the inverse transformation functions $\hat{\varphi}_i=\Phi_i(\hat{\theta}_i)$ and $\hat{\varphi}_j=\Phi_j(\hat{\theta}_j)$, (2) or by projecting the coupling functions onto the domain of the observable phases using the forward transformation functions $q_{ij}(\theta_i,\theta_j) = q_{ij}(\theta_i=\Theta_i(\varphi_i),\theta_i=\Theta_i(\varphi_i))$.
\end{enumerate}

These functions are also included in the code of the examples. For more details we refer to the documentation of the eDCM PC package.  

\section*{(2) Availability}
\vspace{0.5cm}
\section*{Operating system}

\begin{itemize}
\item Windows (\textgreater=XP 32bit, SP2), 
\item Linux (\textgreater=Kernel 2.4.x or 2.6.x and \textgreater= glibc (glibc6) 2.3.4),
\item MacOS (\textgreater=X 10.4.7)
\end{itemize}
Please refer to the system requirements of MATLAB 2007a, which can be found here: \url{https://de.mathworks.com/support/requirements/previous-releases.html}

\section*{Programming language}

The numerical computing language MATLAB.

\section*{Additional system requirements}

Processor Intel Pentium IV and above, Disk space \textgreater=1024MB (500MB MATLAB and 350 MB SPM), RAM \textgreater=1024MB. 

\section*{Dependencies}

eDCM PC needs MATLAB and SPM:
\begin{enumerate}
 \item MATLAB (\textgreater=7.4, R2007a)
 \item Statistical Parametric Mapping (SPM) (SPM12, \textgreater= Update 6225)
\end{enumerate}

\section*{List of contributors}

Please refer to the list of authors.

\section*{Software location:}

{\bf Archive} 

\begin{description}[noitemsep,topsep=0pt]
	\item[Name:] 
	extended Dynamic Causal Modelling for Phase Coupling (eDCM PC)
	
	\item[Persistent identifier:] 
	\url{https://doi.org/10.5281/zenodo.5782819}
	\item[Licence:] 
	Creative Commons Attribution 4.0 International
	\item[Publisher:] 
	Azamat Yeldesbay
	\item[Version published:] 
	Version 1
	\item[Date published:] 
	15/12/2021
\end{description}

{\bf Code repository} 

\begin{description}[noitemsep,topsep=0pt]
	\item[Name:] GitLab
	\item[Persistent identifier:] 
	\url{https://gitlab.com/azayeld/edcmpc}
	\item[Licence:]  
	GPLv3 \url{https://gitlab.com/azayeld/edcmpc/-/blob/master/LICENSE}
	\item[Date published:] 18/02/2019 
\end{description}

\section*{Language}

English

\section*{(3) Reuse potential}
Despite being originally developed to analyze the connectivity between oscillating brain regions, eDCM PC makes no assumption about the origin of the oscillating system. Therefore, the reuse potential of eDCM PC is not restricted to brain signals, but it can also be used in other biological as well as physical and mechanical systems. In general, this software can be used to reconstruct the effective connectivity in any network of oscillating systems. The only requirement is the presence of the phase information extracted from the rhythmic signals for every element of the network. 

The collection of codes of eDCM PC and the testing and example scripts, together with the documentation are uploaded in the GitLab repository \url{https://gitlab.com/azayeld/edcmpc}, where users can clone and modify the code. The modifications can also be included into the package by sending a merge request. Moreover, the GitLab repository has a build-in support mechanism in the form of an issue tracking system.  In the issue tracking system the users of eDCM PC can post their suggestions and problems with the code by opening an issue, which is monitored and will be solved by the authors. 

\section*{Acknowledgements}

We would like to thank Gereon Fink for useful discussions on this project.

\section*{Funding statement}

This work was supported by the Deutsche Forschungsgemeinschaft (DFG, German Research Foundation) – Project-ID 431549029 – SFB 1451.

\section*{Competing interests}

The authors have no competing interests to declare.

%\section*{References}

\vspace{2cm}

%\bibliographystyle{jorsbst}
%\bibliography{libphcon_copy}

\rule{\textwidth}{1pt}

{ \bf Copyright Notice} \\
Authors who publish with this journal agree to the following terms: \\

Authors retain copyright and grant the journal right of first publication with the work simultaneously licensed under a  \url{http://creativecommons.org/licenses/by/3.0/}{Creative Commons Attribution License} that allows others to share the work with an acknowledgement of the work's authorship and initial publication in this journal. \\

Authors are able to enter into separate, additional contractual arrangements for the non-exclusive distribution of the journal's published version of the work (e.g., post it to an institutional repository or publish it in a book), with an acknowledgement of its initial publication in this journal. \\

By submitting this paper you agree to the terms of this Copyright Notice, which will apply to this submission if and when it is published by this journal.

\end{document}